\documentclass[journal,twoside,web]{ieeecolor}
\usepackage{tmi}
\usepackage{amsmath,amssymb,amsfonts}
\usepackage{algorithmic}
\usepackage{graphicx}
\usepackage{textcomp}
\usepackage{epsfig}
\usepackage{multirow}
\usepackage{bm}
\usepackage{changepage}
\usepackage{float}
\usepackage{mathrsfs}
\usepackage{fancyhdr}
\usepackage{booktabs}
\usepackage{array}
\usepackage[ruled,vlined]{algorithm2e}
\usepackage{algorithmic}

\usepackage{hyperref}
\hypersetup{%
	colorlinks=true,
	linkcolor=red,
	anchorcolor=green,
	citecolor=green}

\def\BibTeX{{\rm B\kern-.05em{\sc i\kern-.025em b}\kern-.08em
    T\kern-.1667em\lower.7ex\hbox{E}\kern-.125emX}}
\begin{document}

\title{Viral Pneumonia Screening on Chest X-rays Using Confidence-Aware Anomaly Detection}
\author{Jianpeng~Zhang,
	Yutong~Xie,
    Guansong~Pang,
	Zhibin~Liao,
	Johan~Verjans,
	Wenxing~Li,
	Zongji~Sun,
	Jian~He,
	Yi~Li,
	Chunhua~Shen,
	and~Yong~Xia%
	\thanks{This work was supported in part by the National Natural Science Foundation of China under Grants 61771397 and in part by the Science and Technology Innovation Committee of Shenzhen Municipality, China, under Grants JCYJ20180306171334997. Y. Xie was supported by the Innovation Foundation for Doctor Dissertation of Northwestern Polytechnical University under Grants CX202010.
	({\em J. Zhang and Y. Xie contributed equally to this work.})
	({\em Corresponding authors: C. Shen and Y. Xia})  }%
	\thanks{J. Zhang, Y. Xie, and Y. Xia are with the National Engineering Laboratory for Integrated Aero-Space-Ground-Ocean Big Data Application Technology, School of Computer Science and Engineering, Northwestern Polytechnical University, Xi'an 710072, China; Y. Xia is also with the Research \& Development Institute of Northwestern Polytechnical University in Shenzhen, Shenzhen 518057, China;  
	G. Pang, Z. Liao,  J. Verjans, and C. Shen are with The University of Adelaide, Australia; 
	Y. Li is with the GreyBird Ventures, LLC; W. Li, and Z. Sun are with the JF Healthcare Inc. J. He is with the Department of Radiology, Nanjing Drum Tower Hospital-Affiliated Hospital of Nanjing University Medical School.}
	\thanks{The first two authors' contributions were made when visiting The University of Adelaide. Z. Liao, G. Pang, J. Verjans, C. Shen and their employer received no financial support for the research, authorship, and/or publication of this article. (e-mail: james.zhang@mail.nwpu.edu.cn; xuyongxie@mail.nwpu.edu.cn;  chunhua.shen@adelaide.edu.au; yxia@nwpu.edu.cn)}
}

\maketitle

\begin{abstract}
Clusters of viral pneumonia occurrences over a short period may be a harbinger of an outbreak or pandemic. 
Rapid and accurate detection of viral pneumonia using chest X-rays can be of significant value for large-scale screening and epidemic prevention, particularly when other more sophisticated imaging modalities are not readily accessible. 
However, the emergence of novel mutated viruses causes a substantial dataset shift, which can greatly limit the performance of classification-based approaches.

In this paper, we formulate the task of differentiating viral pneumonia from non-viral pneumonia and healthy controls into a one-class classification-based anomaly detection problem. We therefore propose the confidence-aware anomaly detection (CAAD) model, which consists of a shared feature extractor, an anomaly detection module, and a confidence prediction module. If the anomaly score produced by the anomaly detection module is large enough, or the confidence score estimated by the confidence prediction module is small enough, the input will be accepted as an anomaly case (\textit{i.e.}, viral pneumonia).

The major advantage of our approach over binary classification is that we avoid modeling individual viral pneumonia classes explicitly and treat all known viral pneumonia cases as anomalies to improve the one-class model. 
The proposed model outperforms binary classification models on the clinical X-VIRAL dataset that contains 5,977 viral pneumonia (no COVID-19) cases, 37,393 non-viral pneumonia or healthy cases. Moreover, when directly testing on the X-COVID dataset that contains 106 COVID-19 cases and 107 normal controls without any fine-tuning, our model achieves an AUC of 83.61$\%$ and sensitivity of 71.70$\%$, which is comparable to the performance of radiologists reported in the literature.

\end{abstract}


\section{Introduction}

\begin{figure}[h]
\centering
\includegraphics[height=8.0cm]{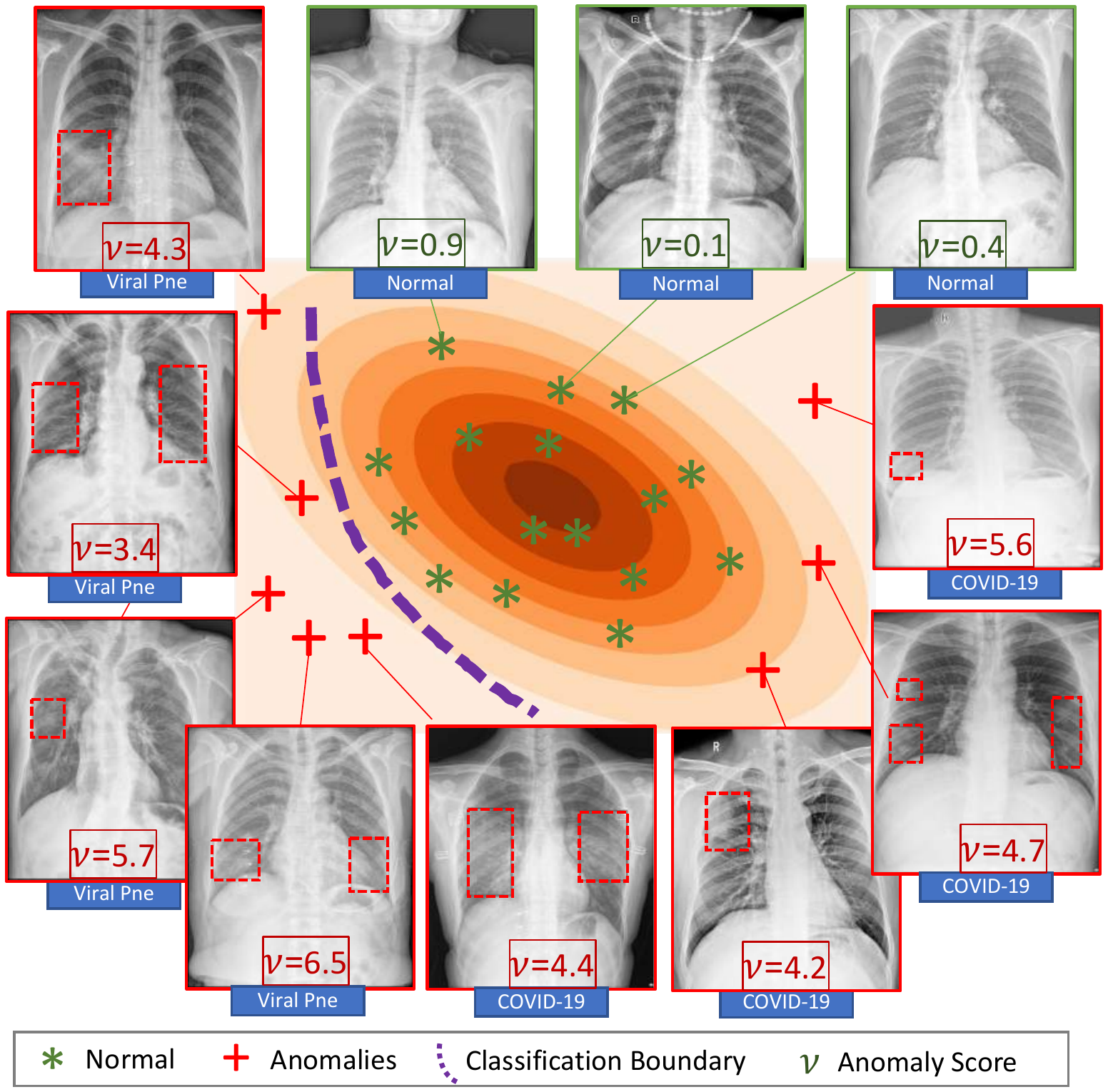}
\caption{An illustration of binary classification vs. anomaly detection in distinguishing viral pneumonia cases (\textit{i.e.}, `Anomalies') from non-viral cases and healthy controls (\textit{i.e.}, `Normal'). Image samples are shown on a two-dimensional contour plot, where inner contour lines indicate large density. The red dotted boxes in viral pneumonia cases are the suspected lesion areas annotated by radiologists. Novel viral pneumonia (\textit{i.e.}, COVID-19) can be either similar to or different from the cases caused by known viruses (denoted as `Viral Pne'). The decision boundary given by a binary classification approach can well separate `Normal' cases and `Viral Pne' cases, but may not able to distinguish `Normal' cases from COVID-19 cases. In contrast, our anomaly detection approach can distinguish both known and novel viral pneumonia from `Normal' cases by assigning the former large anomaly scores and the latter small anomaly scores. 
}
\label{fig:challenge}
\end{figure}

\begin{figure*}[h]
	\centering
\includegraphics[height=9.0cm]{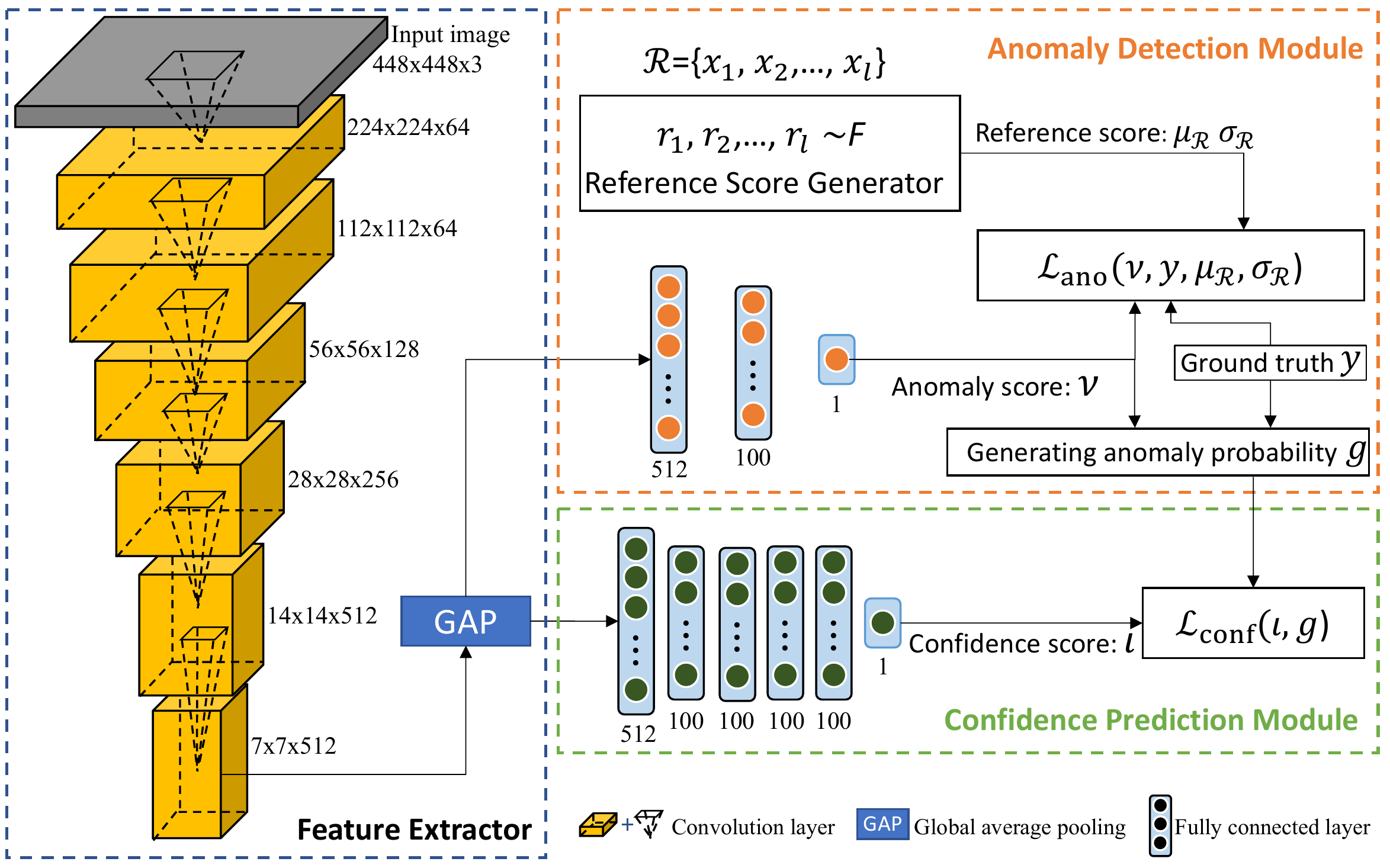}
	\caption{Diagram of the proposed CAAD model. This model is composed of an anomaly detection module and a confidence prediction module, which are designed to predict the anomaly score and confidence score of each input, respectively.}
	\label{fig:framework}
\end{figure*}

\IEEEPARstart{V}{iral} pneumonia is a type of lung infection caused by viruses.
It
causes approximately 30 per cent of pneumonia cases, and is typically mild and healing without treatment. However, COVID-19 has demonstrated to be a much more dangerous virus causing severe pneumonia in a subset of patients, and has rapidly spread throughout the globe within a few months, threatening the health of billions of human beings with significant mortality~\cite{bai2020presumed,chen2020clinical}. 
The clustering aspect of viral pneumonia occurrences, like SARS \cite{li2003angiotensin}, MERS \cite{azhar2014evidence}, and COVID-19, can often suggest a potential outbreak. Therefore, it is desirable to develop an accurate, fast, and cost-effective tool for viral pneumonia detection, which provides the prerequisite for rigorous detection, contact tracing, and isolation of infected subjects in a large district.

Take the recent COVID-19 outbreak for example.
Viral nucleic acid detection using real-time polymerase chain reaction (RT-PCR) is the accepted standard diagnostic method \cite{corman2020detection,lan2020positive}. 
However, the test has suboptimal sensitivity and specificity and many hyperendemic regions and countries are not able to provide sufficient RT-PCR testing for tens of thousands of suspected subjects in a short period of time. Moreover, it would fail to detect the newly evolved coronavirus before extracting the DNA sequence of the new virus from the swab, which may delay the control of the epidemic \cite{watson2020interpreting}. 

Accordingly, medical imaging is considered as a vital technique to assist doctors to evaluate the disease and to optimise prevention and control measures. 
Clinically, chest X-ray is the most commonly used imaging modality in the diagnostic workup of patients with thoracic abnormalities, due to its fast imaging speed, low radiation and low cost \cite{self2013high}. 
In comparison to computed tomography (CT), chest X-ray cannot provide 3D anatomy but is generally considered to be able to differentiate pneumonia although it is probably considered the most challenging plain film to interpret correctly \cite{joarder2009chest}. 
Accurate interpretation is vital to patient management in the acute setting, and to help identify clustering occurrence of COVID-19. Therefore, we aim to develop an automated and accurate viral pneumonia screening technique using chest X-ray as a stopgap for early warning of clusters of an outbreak caused by COVID-19 or a novel coronavirus.

Recent advances in deep learning have led to breakthroughs in many long-standing medical image analysis tasks, such as the detection, staging, and delineation of pathological abnormalities. On the task of chest X-ray interpretation, deep convolutional neural networks (DCNNs) have been constructed to diagnose the most common and important thoracic diseases \cite{luo2020deep,wang2017chestx,rajpurkar2017chexnet,wang2019thorax} and to differentiate between viral and bacterial pneumonia \cite{kermany2018identifying, rajaraman2018visualization}.
In contrast, we attempt to distinguish viral pneumonia from all non-viral pneumonia ones (not merely bacterial pneumonia), aiming to rapidly detect the clusters of viral pneumonia (\textit{e.g.}, COVID-19) caused by a novel virus before an outbreak.
This task, nevertheless, remains challenging due to two intrinsic complexities. First, many different types of viruses can cause pneumonia, including the influenza A/B viruses, respiratory syncytial virus, coronaviruses, herpes simplex, measles, chickenpox, and more seriously, some novel viruses. The complex pathological cues of viral pneumonia cause considerable visual differences on X-ray images (see Fig.~\ref{fig:challenge}), leading to substantial intra-class variance and dataset shift (\textit{e.g.,} newly emerged viral pneumonia cases have highly different lesions from the known viral pneumonia in the training data).
Second, it is hard to collect a large number of positive (\textit{i.e.}, viral pneumonia) samples in the early stage of an outbreak. Hence, the viral versus non-viral pneumonia classification is an extreme class-imbalance problem. 
These two complexities pose significant challenges to the commonly-used binary classification approaches since 1) they normally assume that the training and test data share an identical distribution (\textit{i.e.,} no dataset shift) and there exists small class variance within each class; and 2) they often ignore the class-imbalance problem. Consequently, the classifiers yield poor sensitivity performance. The sensitivity, however, is clinically significant, since it can be much more disastrous to discharge a patient with viral pneumonia than to misdiagnose a healthy control. 
To address both challenges, we advocate the replacement of a classifier by an anomaly detector for this chest X-ray interpretation problem. 
As an one-class classification approach \cite{bellinger2012one}, anomaly detection is not only able to detect dissimilar or even previously unseen anomalies, but also less dependent on labeled anomaly data than binary classification \cite{zhou2017anomaly,pang2018learning,pang2019deep}.

In this paper, we propose a confidence-aware anomaly detection (CAAD) model to distinguish viral pneumonia cases from non-viral pneumonia cases and healthy controls using chest X-rays. 
We reformulate the viral pneumonia screening into a one-class classification-based anomaly detection task, instead of a binary classification one. 
Specifically, we design an anomaly detection module to assign each X-ray image an anomaly score and employ the contrastive loss function to ensure that the scores generated for anomalies (\textit{i.e.}, viral pneumonia) are significantly larger than those for non-viral pneumonia cases and normal controls.
We further introduce an additional confidence prediction module to describe the confidence of the anomaly detection module. 
According to
the confidence level, we re-assign the samples with low confidence as suspected viral pneumonia for further medical tests, which helps reduce false-negative cases and thus
improves 
the sensitivity. 
Both the anomaly detection module and confidence prediction module can be jointly optimized in an end-to-end manner. 
We have evaluated our CAAD model on the X-VIRAL dataset which contains 5,977 positive viral pneumonia subjects (anomalies) and 37,393 negative subjects (non-viral pneumonia and health cases, known as normal controls).
Our proposal achieves the state-of-the-art performance, \textit{i.e.}, 87.57\% AUC, for viral pneumonia screening. 
Even with no exposure to COVID-19 cases during training, our CAAD model shows superior performance for the COVID-19 screening purpose, achieving an AUC of $83.61\%$ and a sensitivity of $71.70\%$ on our additional unseen X-COVID dataset with 106 confirmed and 107 normal subjects. This is comparable to the performance of radiologists reported in the literature \cite{wong2020frequency}.

The main contributions of this paper are summarized as follows.
\begin{itemize}
    \item  We formulate  viral pneumonia screening into an anomaly detection problem and propose the CAAD model to solve it, which is able to detect viral pneumonia caused by novel viruses and is less dependent on labeled viral pneumonia data than classification models.
    
    \item 
    We propose to predict failures of anomaly detection by modeling its confidence level so as to further improve the screening sensitivity.
    
    \item 
    Our experimental results demonstrate the effectiveness and strong generalizability of our model in viral pneumonia screening and the potential on epidemic prevention and control.

\end{itemize}

\section{Related work}

\subsection{Chest X-ray for pulmonary disease screening}
Chest X-ray is one of the most commonly used imaging modalities for visualizing and quantifying the structural and functional consequences of thoracic diseases, providing high-resolution pictures of disease progression and therapy response.
Magree \textit{et al.}\  \cite{magree2005chest} documented the incidence of pneumonia confirmed with X-ray imaging and demonstrated a high incidence, which guided the later prevention and treatment of vaccine. 
Jacobi \textit{et al.}\  \cite{jacobi2020portable} described the most common manifestations and patterns of lung abnormality on chest X-ray in COVID-19 and suggested that the medical community can frequently rely on portable chest X-ray due to its widespread availability.
Wong \textit{et al.}\  \cite{wong2020frequency} demonstrated that the common CT findings of bilateral involvement, peripheral distribution, and lower zone dominance can also be appreciated on chest X-ray, which shows the potential of using chest X-ray as a tool for COVID-19 identification. 
Borghesi \textit{et al.}\  \cite{borghesi2020covid} presented an experimental chest X-ray scoring system and applied it to hospitalized patients with COVID-19 to quantify and monitor the severity and progression of COVID-19. 
Different from these studies, we focus on the viral pneumonia screening and aim to develop a fast and accurate algorithm to differentiate viral pneumonia from non-viral pneumonia and normal controls for the prevention and control of a possible outbreak.

\subsection{Deep learning for chest X-ray interpretation}

To improve efficiency and ease the burden of radiologists, researchers gradually adapt the recent advances of deep learning to interpret chest X-ray images.
For computer-aided diagnosis of 14 common thoracic diseases, Wang \textit{et al.} \cite{wang2017chestx} proposed a weakly-supervised classification and localization framework, Rajpurkar \textit{et al.} \cite{rajpurkar2017chexnet} constructed a 121-layer dense convolutional neural network that can perform the task at a level exceeding practicing radiologists, and Wang \textit{et al.} \cite{wang2019thorax} introduced an attention mechanism to help the model focus on the lesion area and thus further improved the diagnosis performance.
Besides, many attempts \cite{kermany2018identifying, rajaraman2018visualization} have been made to investigate DCNN-based classification models for pneumonia detection and the differentiation between viral and bacterial pneumonia, aiming to facilitate rapid referrals for children who need urgent intervention. 
In these studies, these diagnostic tasks are formulated as classification problems, which are usually solved based on the intra-class similarity and inter-class dissimilarity of pathological patterns. Such classification models may fail to distinguish viral and non-viral pneumonia since the category of viral pneumonia contains cases with highly variable visual appearances.

\subsection{Failure prediction}
Despite their success, deep learning models still make mistakes, particularly when applied to real-world applications. To avoid the decision risk caused by the inherent defects of deep learning models, failure prediction is of great necessity.
Hendrycks \textit{et al.} \cite{hendrycks2016baseline} proposed to detect failures and out-of-distribution examples in neural networks via the prediction/maximum class probability method. However, it is hard to distinguish the failures if they are misclassified with a high probability. To address this issue, Corbiere \textit{et al.} \cite{corbiere2019addressing} considered the true class probability as a suitable confidence criterion for failure estimation. 
Xie \textit{et al.} \cite{xie2019deep,xie2020sesv} proposed a deep segmentation-emendation model for gland instance segmentation, in which an emendation network is designed to predict the inconsistency between the ground truth masks and pixel-wise predictions of segmentation network, and the failure predictions made by an emendation network can be utilized to refine the segmentation result.
Inspired by these works, we attempt to predict failures of anomaly detection, which improves not only the detection performance but also the diagnosis credibility of our model.

\subsection{Deep anomaly detection}
Anomaly detection is the task of identifying unusual samples from the majority of the data \cite{chandola2009anomaly}. 
Traditional anomaly detection mainly focuses on the kernel-based one-class classification. Typical methods include the One-Class SVM (OC-SVM) \cite{scholkopf2001estimating} and Support Vector Data Description (SVDD) \cite{scholkopf2001estimating}, which attempt to separate the anomalies from normal controls using a hyperplane. 
These methods, however, suffer from a poor computational scalability and the curse of dimensionality~\cite{ruff2018deep}. 
By contrast, deep learning specializes in automatically learning feature representations from large-scale data. Many research efforts have been devoted to transfer the advantages of deep learning to anomaly detection. 
Ruff \textit{et al.} \cite{ruff2018deep} presented a deep SVDD model, which uses a deep neural network to minimize the volume of a hypersphere. 
Schlegl \textit{et al.} \cite{schlegl2019f} proposed a fast unsupervised anomaly detection framework with generative adversarial networks (f-AnoGAN), which is able to detect the unseen anomalies from health subjects after being trained on healthy OCT images. 
Although superior to traditional methods, these deep anomaly detection models, also known as novelty detection, employ unsupervised deep learning techniques, such as autoencoders and generative adversarial networks, to characterize the normal class, without using the information provided by anomalies \cite{pang2019deep,siddiqui2018feedback,pang2018learning}.
Differently, we leverage the anomaly data, $i.e.$, clinically available viral pneumonia, in the training procedure, using them as the prior knowledge to reinforce the one-class model. Readers can refer to \cite{pang2020deep} for a detailed review of more deep anomaly detection methods.

\section{Methods}

The proposed CCAD model is composed of an anomaly detection network and a confidence prediction network (see Fig.~\ref{fig:framework}). Both networks share a feature extractor. 
Given an input chest X-ray image $\textbf{x}$, the anomaly detection network aims to learn an anomaly scoring function $\varphi: \textbf{x}\rightarrow \mathbb{R}$. For any two inputs $\textbf{x}_i$ and $\textbf{x}_j$, we have $\varphi (\textbf{x}_i) > \varphi (\textbf{x}_j)$ if $\textbf{x}_i$ is abnormal and $\textbf{x}_j$ is normal. The confidence prediction network targets at approximating a confidence scoring function $\zeta: \textbf{x} \rightarrow [0, 1]$, where 1 indicates highest model confidence and 0 indicates the opposite.
In the inference mode, if the anomaly score is larger than $T_{ano}$ or the confidence score is less than $T_{conf}$, we accept the input as an anomaly case (\textit{i.e.}, viral pneumonia). We now delve into each part of our model.

\subsection{Feature extractor}

\begin{table}[t]
\caption{Architecture of our feature extractor. S: stage, L: number of stacked layers, $W$/$H$/$C$: weight/height/channel, GAP: Global average pooling}
\vspace{+0.2cm}
\label{tab:effinet}
\centering
\begin{tabular}{c|c|c|c}
\hline
\textbf{S} & \textbf{Operator}   & \textbf{Input $\rightarrow$ Output ($W\times H \times C$)} & \textbf{L} \\ \hline
\multirow{2}{*}{1} & Conv3x3 & $448 \times 448 \times 3\rightarrow 224 \times 224 \times32$ & 1 \\ \cline{2-4} 
  & MBConv1, k3x3 & $224 \times 224 \times 32\rightarrow 224 \times 224\times 16$     & 1      \\ \hline
2 & MBConv6, k3x3 & $224 \times 224 \times 16\rightarrow 112 \times 112 \times 24$     & 2      \\ \hline
3 & MBConv6, k5x5 & $112 \times 112 \times 24\rightarrow 56 \times56 \times 40$       & 2      \\ \hline
\multirow{2}{*}{4} & MBConv6, k3x3 & $56 \times 56 \times 40\rightarrow 28 \times 28 \times 80$    & 3  \\ \cline{2-4} 
                   & MBConv6, k5x5 & $28 \times28 \times 80\rightarrow 28 \times 28 \times 112$   & 3  \\ \hline
\multirow{2}{*}{5} & MBConv6, k5x5 & $28 \times 28 \times 112\rightarrow 14 \times 14 \times 192$ & 4 \\ \cline{2-4} 
  & MBConv6, k3x3 & $14 \times 14 \times 192\rightarrow 14 \times 14 \times 320$       & 1      \\ \hline
6 & Conv3x3       & $14 \times 14 \times320 \rightarrow 7 \times 7 \times 320$         & 1      \\ \hline
  & GAP           & $7 \times 7 \times 320 \rightarrow 1 \times 1 \times 320$           & 1      \\ \hline
\end{tabular}
\end{table}

Although a DCNN with any architecture can be embedded in our CAAD model as the feature extractor, we choose the state-of-the-art EfficientNet \cite{tan2019efficientnet} with the B0 architecture pretrained on ImageNet \cite{deng2009imagenet}, due to the trade-off between the performance and complexity. This network is mainly composed of mobile inverted bottleneck (MBConv) blocks \cite{sandler2018mobilenetv2, tan2019mnasnet} with squeeze-and-excitation module \cite{hu2018squeeze}. The six stages of layer-by-layer convolution operations are represented by yellow rectangles in Figure~\ref{fig:framework}, and the architecture details are listed in Table~\ref{tab:effinet}.
For each input chest X-ray image $\textbf{x}$, it is first processed by several MBConv blocks, and then transformed into a $d$-dimensional ($d$ equals the number of channels in the last convolution layer) feature vector by a global average pooling layer. We denote the parameters of feature extractor as $\bm{\theta}$.

\subsection{Anomaly detection network}

The anomaly detection network is composed of the feature extractor and anomaly detection module that is a multi-layer perceptron with three 100-neuron hidden layers and an one-neuron output layer.
It aims to generate an anomaly score for each input image $\textbf{x}$, formulated as 
\begin{equation}
\nu = \varphi (\textbf{x}; \bm{\theta}, \bm{\alpha})
\label{eq.forward_ano}
\end{equation}
where $\bm{\alpha}$ is the trainable parameter of the anomaly detection module.

Extensive results in \cite{kriegel2011interpreting} show that Gaussian distribution fits the anomaly scores very well in a range of datasets.
To guide the learning procedure of the anomaly detection module, we compute another scalar score as a reference. 
We randomly sample $l$ scalar values from an univariate Gaussian distribution, \textit{i.e.}, $r_1, r_2, ...,r_l \sim {\cal N}(\mu, \sigma^2)$, and define a reference score as $\mu_R = \frac{1}{l} \sum_{i=1}^{l}r_i$ and $\sigma_R^2 = \frac{1}{l} \sum_{i=1}^{l} (r_i - \mu_R)^2$. Following \cite{pang2019deep}, we set $\mu=0$, $\sigma=1$, and $l=5,000$.
With the obtained anomaly score and reference score, we employ the following contrastive loss \cite{hadsell2006dimensionality, hinton2020} to optimize the anomaly detection module
\begin{equation}
\begin{split}
\mathcal{L}_{ano}  (\nu, y, \mu_R, \sigma_R) = (1-y) \left| 
\frac{\nu-\mu_R}{\sigma_R}
\right|
\\ + y \, \max \Big( 0, {\tt margin} -\frac{\nu-\mu_R}{\sigma_R}
\Big)
\end{split}
\label{eq.loss_ano}
\end{equation}
where $\sigma_R$ is the standard deviation of the anomaly scores of randomly selected $l$ normal data, $y$ is the ground truth label, \textit{i.e.}, $y=0$ indicates that the input is a negative case and $y=1$  indicates that the input is a positive case. Besides, $\tt margin$ represents the Z-score confidence interval parameter, which is empirically set to 5 for this study.

Different from binary classification where performance can be largely degraded when there is imbalanced class distribution, our one-class classification-based anomaly detection network is inherently resilient to the class imbalance. The reason is that the anomaly detection network aims to learn a one-class description model from the large-scale negative data (\textit{i.e.,} non-viral pneumonia). By doing so, it avoids modeling the positive class with the limited amount of labeled data; the limited positive data is used instead to reinforce the one-class modeling to achieve tighter one-class description. 

\subsection{Confidence prediction network}
\label{sec.confidence_prediction}

The current approach of anomaly detection does not have an error correction mechanism. However, we observe that the model does produce false predictions. To alleviate this issue, we follow the work of failure prediction in image classification and segmentation \cite{corbiere2019addressing, xie2019deep}, and make the shift in thinking that we can predict the failures of anomaly detection. Hence, we propose a confidence prediction network to learn a confidence score for each input, which reflects the confidence of the anomaly score estimated by our model. 

\subsubsection{Confidence criterion for anomaly detection}

The predicted anomaly score $\nu \in \mathbb{R}$ explicitly describes the abnormality degree of a given image, varying from very confirmed positive cases, \textit{i.e.}, viral pneumonia ($\nu >={\tt margin}$), to confirmed negative cases, \textit{i.e.}, non-viral pneumonia or healthy patients ($\nu \approx 0$). However, it is difficult to describe the abnormality degree in the form of probability, which is important for confidence prediction. Hence, we propose to employ the probability density function (PDF) to estimate the prediction probability. 
In order to normalize $\nu$, we introduce the Gaussian PDF as
\begin{equation}
{\rm PDF}(\nu) = \frac{1}{\sigma \sqrt{2\pi}}\exp(- \frac{(\nu - \mu)^2}{2\sigma}),
\end{equation}
We then approximate the prediction probability of anomaly detection by the normalized PDF, where PDF is commonly regarded as relative probability and the normalization scales the values into [0, 1], similarly to the use of sigmoid, hyperbolic tangent, or softmax functions in deep learning. The approximated prediction probability is expressed as:
\begin{equation}
\begin{split}
{\rm prob} &= \frac{{\rm PDF}(\nu)}{ \max({\rm PDF}) }  \\
&=\exp(-\frac{(\nu-\mu)^2}{2\sigma}).
\end{split}
\label{eq.normPDF}
\end{equation}
Even if the scores estimated by the model is not Gaussian, it will not affect the computation of the prediction probability.
However, it is hard to use such a prediction probability to distinguish failure predictions from successful ones (discussed in Section~\ref{sec.failure_prediction}). 
To further address this issue, we propose the anomaly probability, formulated as:
\begin{equation}
g =\left\{
\begin{array}{l l}
{{\rm prob}}            & \rm if \ y = 0 \\
1 - {\rm prob}         & \rm if \ y = 1
\end{array}
\label{eq.AP}
\right..\end{equation}
where $g \in [0, 1]$ is the confidence criterion for distinguishing successful and erroneous predictions. Intuitively, a robust model should successfully predict the true labels of input cases with a high confidence, whereas should have low confidence when making erroneous predictions. Therefore, in Eq.~\eqref{eq.AP}, $g$ is close to 0 when the anomaly detection module fails to predict the true labels of given image and close to 1 when the true labels are successfully predicted.

\subsubsection{Confidence prediction network}
Our confidence prediction network is also built upon the shared feature extractor, and particularly contains the confidence prediction module with four 100-neuron hidden layers (see Fig.~\ref{fig:framework}) for a strong prediction ability, as done in \cite{corbiere2019addressing}. The forward computation of the confidence prediction network can be formally expressed as
\begin{equation}
\iota = \zeta (\textbf{x}; \bm{\theta}, \bm{\beta})
\label{eq.forward_conf}
\end{equation}
where $\iota$ is the confidence score of the corresponding anomaly detection, and $\bm{\beta}$ represents the ensemble of parameters of this module. 

Since the confidence score $\iota$ takes a value from the range $[0, 1]$, we formulate confidence prediction as a regression task and employ the standard $l2$ loss to optimize the confidence prediction module. 
\begin{equation}
\mathcal{L}_{conf} (\iota, g) = |\iota - g|^2
\label{eq.loss_conf}
\end{equation}

\begin{algorithm}[t]
\footnotesize
\KwIn{$\mathfrak{D} =\{(\textbf{x}_i, \textbf{y}_i)\}_{i=1}^{N}$ - training data and labeled ground truth. Initialize $\bm{\theta}$ of feature extractor with the pretrained weights in ImageNet. Randomly initialize $\bm{\alpha}$ and $\bm{\beta}$.}
\KwOut{Anomaly detection network $\varphi $, and Confidence prediction network $\zeta $.}
\caption{Training CAAD model}
\begin{algorithmic}[1]
\STATE $-$ \textbf{Step1}: \textit{Training anomaly detection network }
\WHILE {\textit{not converge}}
\STATE Randomly sample a batch with $m$ samples %
with
half positive cases and half negative cases
\STATE Randomly sample $l$ scalar values from Gaussian distribution to compute $\mu_R, \sigma_R$ as the reference score
\STATE Compute the anomaly score $\nu_i$ via Eq.~\eqref{eq.forward_ano} for each sample $\textbf{x}_{i}$
\STATE Compute the anomaly detection loss 
\[
\frac{1}{m} \sum_{i=1}^{m} \mathcal{L}_{ano}  (\nu_i, y_{i}, \mu_R, \sigma_R)
\]
\STATE Update parameters $\bm{\theta}$ and $\bm{\alpha}$ by using back-propagation
\ENDWHILE 
\STATE $-$ \textbf{Step2}: \textit{Training confidence prediction network }%
\STATE Fix the feature extractor $\bm{\theta}$ and anomaly detection module $\bm{\alpha}$
\WHILE {\textit{not converge}}
\STATE Randomly sample a batch with $m$ samples
\STATE Compute the confidence score $\iota_i$ via Eq.~\eqref{eq.forward_conf} for each sample $\textbf{x}_{i}$
\STATE Generate the anomaly probability $g_i$ for each sample
\STATE Compute the confidence prediction loss $\frac{1}{m} \sum_{i=1}^{m} \mathcal{L}_{conf}  (\iota_i, g_i)$
\STATE Update parameters $\bm{\beta}$ by using back-propagation algorithm.
\ENDWHILE 
\STATE $ - $ \textbf{Step3}: \textit{Joint training in an end-to-end manner} %
\WHILE {\textit{not converge}}
\STATE Randomly sample a batch with $m$ samples
\STATE Compute $\mu_R, \sigma_R$
\STATE Compute $\nu_i$ and $\iota_i$, then generate $g_i$ according to Eq.~\eqref{eq.AP}
\STATE Update parameters $\bm{\theta}$, $\bm{\alpha}$, and $\bm{\beta}$ by using back-propagation in an end-to-end manner. 
\ENDWHILE
\end{algorithmic}
\label{algorithm1}
\end{algorithm}

\begin{table*}[t]
\caption{%
Comparison of 
results of binary classification models and the anomaly detection model on the X-VIRAL dataset.}
\vspace{+0.2cm}
\label{tab:comp2cls}
\centering
\begin{tabular}{c|c|c|c|c|c|c}
\toprule[1pt]
\multicolumn{2}{c|}{Mode}          & \multirow{2}{*}{Feature extractor} & \multicolumn{4}{c}{Performance \%}                                  \\ \cline{1-2} \cline{4-7} 
Binary classification & Anomaly detection &                                    & Accuracy       & Sensitivity    & Specificity    & AUC            \\ \midrule[1pt]
\checkmark     &                   & ResNet                             & 78.52          & 78.28          & 78.56          & 86.24          \\ \hline
               & \checkmark     & ResNet                             & 80.04          & 84.44          & 79.34          & 87.18          \\ \hline
\checkmark     &                   & EfficientNet                       & 78.71          & 79.09          & 78.65          & 86.30          \\ \hline
               & \checkmark     & EfficientNet                       & \textbf{80.65} & \textbf{85.51} & \textbf{79.87} & \textbf{87.42} \\ \bottomrule[1pt]
\end{tabular}
\end{table*}

\subsection{Training and inference}

We resize each training image to a fixed size of $512\times512$ pixels and applied several data augmentation strategies, including random cropping patches of size $448\times448$ and zooming (90\%$\sim$110\%) and horizontally flipping cropped patches, to alleviate overfitting on the it. Then, the proposed CAAD model is trained in three steps, Following \cite{corbiere2019addressing}, the proposed CAAD model was trained in three steps.
First, we trained the anomaly detection network, which is the combination of the anomaly detection module and shared feature extractor, via minimizing the loss $\mathcal{L}_{ano}$ using the vanilla stochastic gradient descent (SGD) algorithm.
Second, we fixed the feature extractor and trained the confidence prediction network, which is the combination of the confidence prediction module and shared feature extractor, via minimizing the loss $\mathcal{L}_{conf}$ using the Adam algorithm.
In both steps, we set the max epoch to 20, the batch size to 40, and the initial learning rate to $5\times 10^{-4}$, which decays linearly to $10^{-6}$ during the entire training process.
Third, we fine-tuned the entire model in an end-to-end manner via minimizing the combination of $\mathcal{L}_{ano}$ and $\mathcal{L}_{conf}$ with the Adam optimizer. Here, the weight ratio of $\mathcal{L}_{ano}$ and $\mathcal{L}_{conf}$ was set to $1:1$.
In this step, we set the max epoch to 10, the batch size to 40, and the initial learning rate to $10^{-7}$, aiming to avoid deviating too much from the original anomaly detection scores. 
Note that the shared feature extractor was pretrained on ImageNet, and the anomaly detection module and confidence prediction module were randomly initialized.

In the inference stage, we input a test image into the well-trained model and generate a scalar anomaly score $\nu$ and a confidence score $\iota$ via the forward propagation. For the anomaly detection, we assume $g=0.5$ in Eq.~\eqref{eq.AP} as the boundary point. According to Eq.~\eqref{eq.normPDF} and Eq.~\eqref{eq.AP}, we can compute the corresponding boundary anomaly score $\nu \approx 1.18$. Therefore, we set $T_{ano}=1.18$ as the threshold to distinguish the abnormal cases and normal cases, \textit{i.e.}, detected as abnormal if $\nu \geq T_{ano}$ or otherwise. 
As for the confidence prediction, we empirically set $T_{conf}=0.9$ as the threshold to correct the erroneous predictions with low confidence. In practice, we only re-label the predictions, recognized as normal or abnormal, with low confidence, \textit{i.e.}, $\iota < T_{conf}$, as abnormal cases to achieve the high sensitivity, which is significant in clinical study. Therefore, the final diagnosis made by our CAAD model is formulated as:  
\\
\begin{equation}
diag =\left\{
\begin{array}{l l}
1 & if \; \nu \geq T_{ano} \; {\rm or} \; \iota < T_{conf} \\
0 & if \; \nu <T_{ano} \; {\rm and} \; \iota \geq T_{conf}
\end{array}
\right..\end{equation}

Specifically, if either condition is met (\textit{i.e.}, the anomaly score is larger $T_{ano}$ or the confidence score is less than $T_{conf}$), our model gives a $1: {\rm POSITIVE}$ diagnosis and recommend to be further examined by radiologists; otherwise our model gives a $0: {\rm NEGATIVE}$ diagnosis.

\subsection{Performance metrics}

For this study, the diagnostic performance of an algorithm is quantitatively assessed by the area under the receiver operator curve (AUC), sensitivity, specificity, and, accuracy. AUC reflects the probability that a recognition model ranks a randomly chosen positive instance higher than a randomly chosen negative case. It is the most commonly used metric to evaluate the overall classification performance. The sensitivity and specificity give the proportion of positives and negatives that were correctly identified, respectively. Accuracy gives the percentage of correctly classified cases, including both positive and negative ones.

\section{Datasets}

Two in-house X-ray image datasets, X-VIRAL and X-COVID, were used for this study. 
The X-VIRAL dataset contains 5,977 viral pneumonia cases, 18,619 non-viral pneumonia cases, and 18,774 healthy controls (\textit{i.e.}, 5977 positive and 37,393 negative cases) collected from 390 township hospitals through a telemedicine platform of JF Healthcare during 2019. Each X-ray image has a high resolution, varying from 1000 to 3000 for height and weight, and was annotated by one of three board-certified radiologists. 
Note that all viral pneumonia cases were collected before the COVID-19 outbreak, and hence do not contain any COVID-19 cases.
The X-COVID dataset was collected from 6 institutions during March 2020. It consists of 106 confirmed COVID-19 cases and 107 normal controls.
Besides, a public COVID-19 dataset\footnote{\url{  https://github.com/ieee8023/covid-chestxray-dataset}} \cite{cohen2020covidProspective}, called Open-COVID, was used for external validation. This dataset contains the X-ray images of 493 confirmed COVID-19 cases, 16 confirmed SARS \cite{li2003angiotensin} cases, and 10 confirmed MERS \cite{azhar2014evidence} cases.

\section{Experiments and results}

\begin{figure}[t]
\centering
\includegraphics[height=10.8cm]{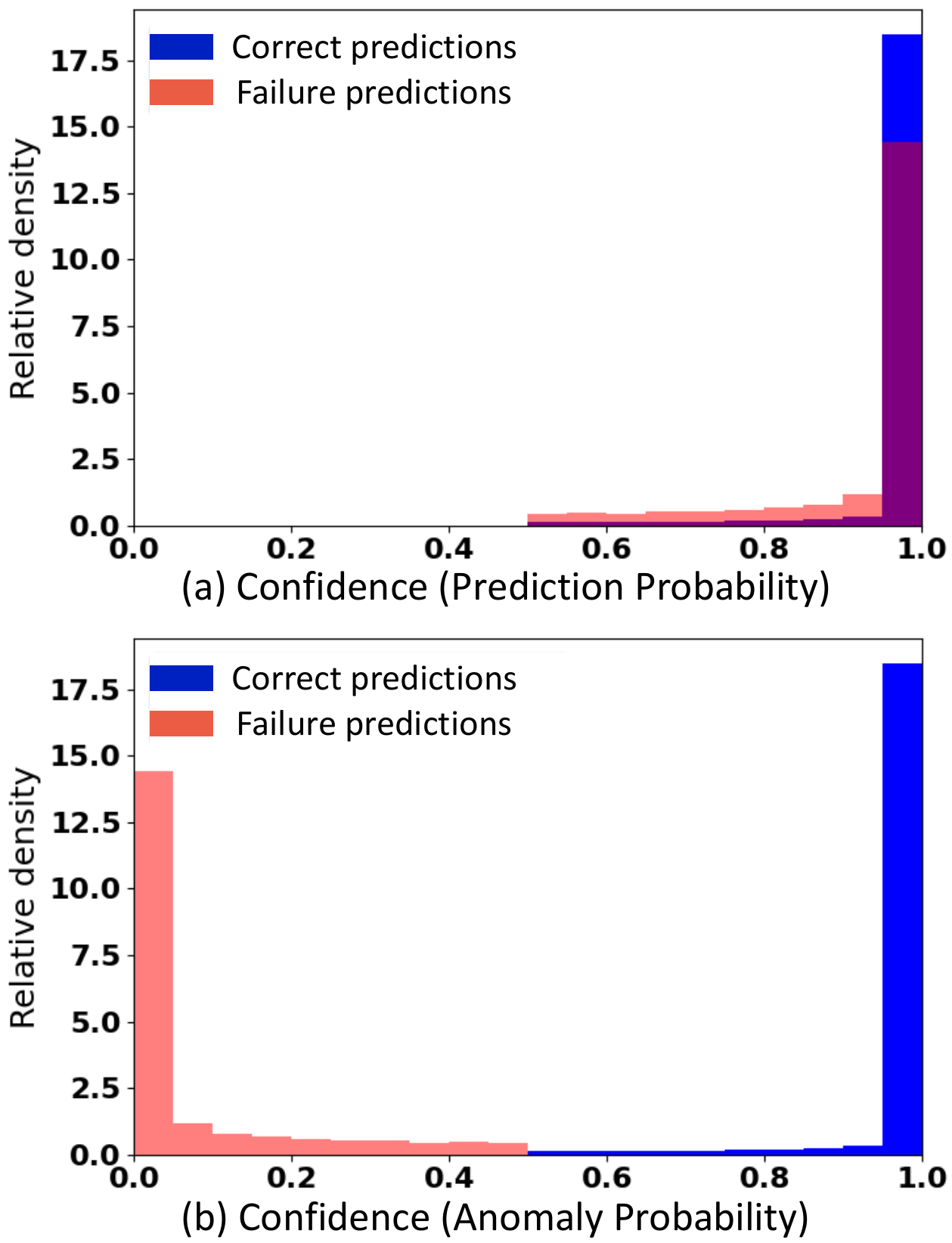}
\caption{Comparison of confidence learning based on (a) the prediction probability and (b) anomaly probability for failure prediction on the validation set of X-VIRAL.}
\label{fig:TCP}
\end{figure}

\begin{figure*}[t]
\centering
\includegraphics[height=7.2cm]{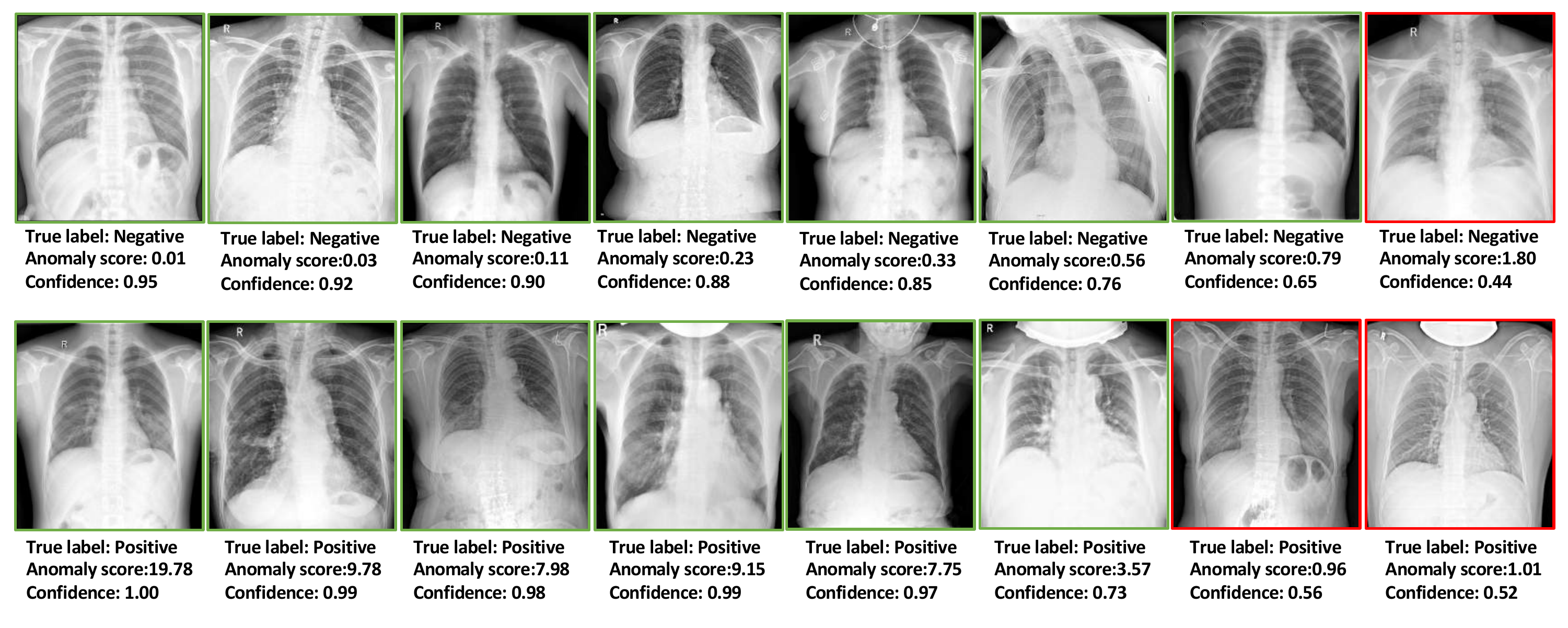}
\caption{A set of 16 chest x-ray images from the X-VIRAL validation set. The predicted anomaly score and confidence score are displayed beneath each image. Positive cases are shown in the top row, and negative cases are shown in the bottom row. Totally, 13 cases (marked with green boxes) were corrected diagnosed by our CAAD model, and three (marked with red boxes) were wrongly diagnosed.}
\label{fig:Failure_Prediction}
\end{figure*}

\subsection{Anomaly detection vs. binary classification}

We first compared anomaly detection models to binary classification models on the X-VIRAL dataset using the five-fold cross-validation. The feature extractor used in all models is either the 18-layer ResNet \cite{he2016deep} or EfficientNet-B0 \cite{tan2019efficientnet}, both being pre-trained on ImageNet. \noindent The binary classification models were optimized by minimizing the cross-entropy loss function. To indulge a fair comparison, we also employed the balanced sampling strategy to keep the balance of both positive and negative categories during the training. The obtained performance metrics were listed in Table~\ref{tab:comp2cls}.
It shows that the binary classification using ResNet achieves the baseline performance, \textit{i.e.}, an accuracy of 78.52\%, a sensitivity of 78.28\%, a specificity of 78.56\%, and an AUC of 86.24\%, which can be improved a little bit by introducing a stronger feature extractor, \textit{i.e.}, EfficientNet-B0.
By contrast, an anomaly detection model always outperforms (particularly in terms of sensitivity) the corresponding binary classification model. In this experiment, the anomaly detection model using EfficientNet-B0 achieves the highest accuracy of 80.65\%, highest specificity of 79.87\%, and highest AUC of 87.42\%, and also remarkably improves the sensitivity from 79.09\% to 85.51\%. 
The results suggest that anomaly detection has distinct advantages over binary classification in distinguishing viral pneumonia cases from non-viral pneumonia cases and healthy controls, especially with an extraordinary ability to detect positive cases as evidenced by a high sensitivity, which is particularly useful for viral pneumonia screening in clinical practice.

\subsection{Confidence learning for failure prediction}
\label{sec.failure_prediction}
To analyze the effectiveness of confidence learning using either the anomaly probability or prediction probability, we compared the distribution of prediction probability and anomaly probability obtained on the validation set of X-VIRAL in Fig.~\ref{fig:TCP}.
It is apparent that the prediction probability distributions of correct predictions and false predictions have a lot of overlap, which hinders the confidence prediction module from accurately distinguishing them. By contrast, the proposed anomaly probability can successfully separate false predictions from correct ones.

In Fig.~\ref{fig:TCP}(b) and Eq.~\eqref{eq.AP}, we observed that, if (1) the predicted anomaly score for negative cases is close to 0 or (2) the predicted anomaly score for positive cases is larger than $T_{ano}$, our CAAD model has high confidence; otherwise low confidence. Such conjecture was confirmed with the results given in Fig.~\ref{fig:Failure_Prediction}, which shows 16 chest X-ray images from the X-VIRAL validation set, each being equipped with the ground truth label and the anomaly scores and confidence scores predicted by our CAAD model.
For each negative case in the top row, it shows that if the predicted anomaly scores are very low, close to 0, the confidence score is close to 1. 
Similarly, for each positive case in the bottom row, it shows that if the predicted anomaly score is larger than $T_{conf}$, the confidence score is still close to 1. 
In contrast, the confidence becomes very low if a case is wrongly diagnosed, as those marked with red bounding boxes. These results demonstrate the effectiveness of using the proposed anomaly probability to learn the confidence for failure prediction.

\begin{figure*}[h]
\centering
\includegraphics[height=11.0cm]{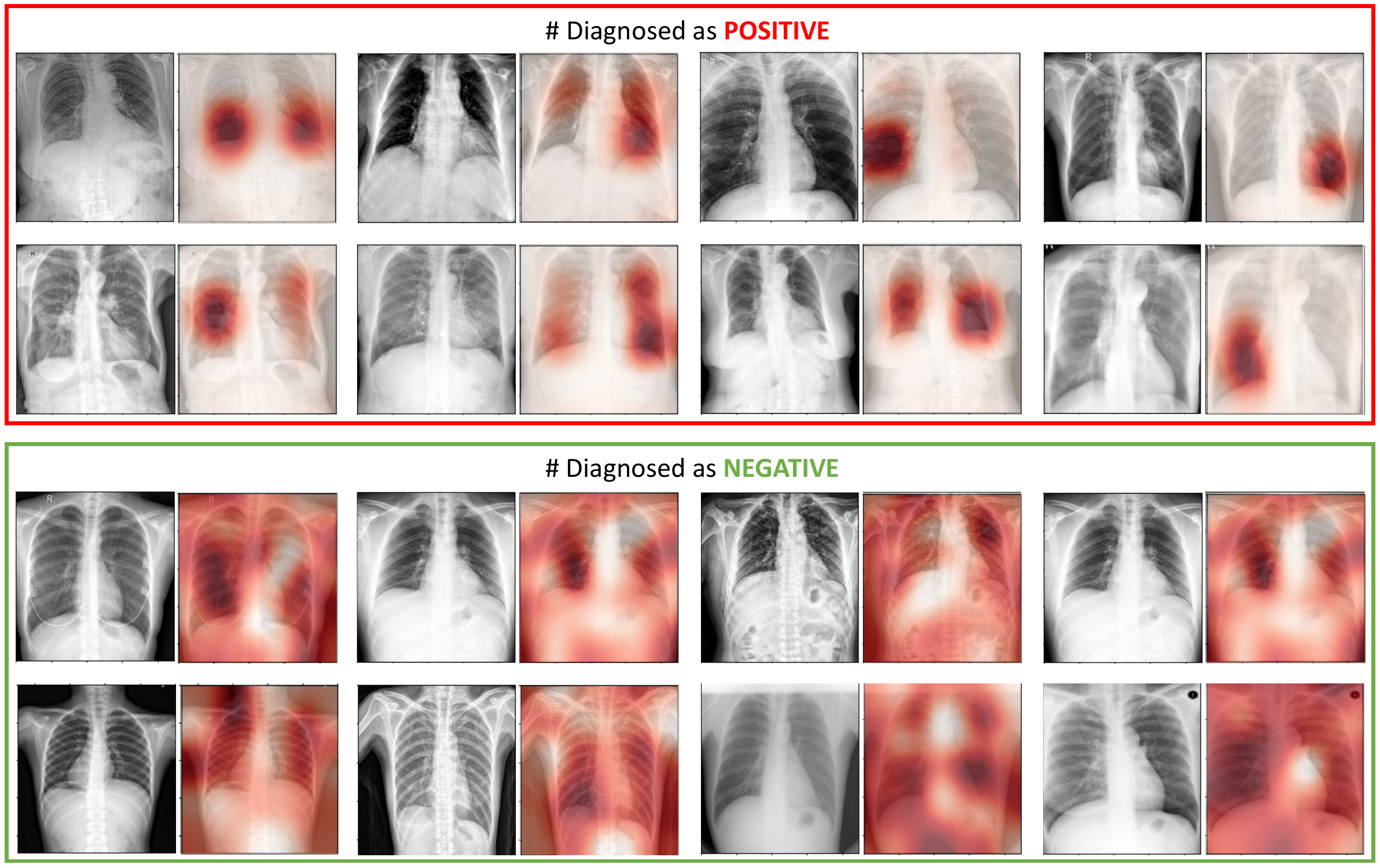}
\caption{Visualization of 16 chest X-ray images from the X-VIRAL validation set and their Grad-CAMs. The images in two top rows are diagnosed as positive by our CAAD model, while the images in two bottom rows are diagnosed as negative.}
\label{fig:CAM}
\end{figure*}

\begin{table}[t]
\caption{Performance of the AnoDet model (without confidence prediction) and our CAAD model (with variable confidence threshold $T_{conf}$) on the X-VIRAL dataset. Note that we set the same threshold $T_{ano}=1.18$ for both models.}
\vspace{+0.2cm}
\label{tab:comp_confi}
\centering
\begin{tabular}{c|c|c|c|c|c}
\toprule[1pt]
Methods & $T_{conf}$    & Accuracy & Sensitivity & Specificity & AUC \\ \midrule[1pt]
AnoDet & /    & 80.65	&85.51	&79.87	&87.42 \\ \hline
\multirow{6}{*}{CAAD} & 0.5  & 80.33	&85.88	&79.44  & \multirow{6}{*}{87.57} \\ \cline{2-5}
 & 0.6  & 79.47	&87.02	&78.27  & \\ \cline{2-5}
 & 0.7  & 78.48	&88.05	&76.95  & \\ \cline{2-5}
 & 0.8  & 76.79	&89.63	&74.74  & \\ \cline{2-5}
 & 0.9  & 71.21	&93.01	&67.72  & \\ \cline{2-5}
 & 0.95 & 46.44	&97.69	&38.25  & \\ \bottomrule[1pt]
\end{tabular}
\end{table}

\subsection{Importance of confidence prediction}

To evaluate the effectiveness of confidence prediction, we compared the anomaly detection network (denoted by AnoDet) with our CAAD model, in which the confidence threshold $T_{conf}$ ranges from 0.5 to 0.95. Note that the CAAD model can be treated as a combination of AnoDet with a confidence prediction module.
The results obtained on the X-VIRAL dataset are shown in Table~\ref{tab:comp_confi}.
It reveals that
(1) using confidence prediction leads to a slightly improved AUC of 87.57\%, improved sensitivity, and deteriorated accuracy and specificity;
(2) when setting the confidence threshold $T_{conf}$ to 0.5, the performance of our model is comparable to that of AnoDet; and
(3) with the increase of the confidence threshold $T_{conf}$, the deterioration of accuracy and specificity becomes severe and the improvement of sensitivity becomes substantial.
Specifically, when setting the confidence threshold $T_{conf}$ to 0.9, our CAAD model can boost the sensitivity from 85.51\% to 93.01\% while having a specificity of 67.72\%. 
The improvement in the sensitivity suggests the advantage of using confidence prediction in our model for the clinical screening of viral pneumonia.
It should noted that, as mentioned in Section~\ref{sec.confidence_prediction}, we propose the confidence prediction module to distinguish the successful predictions with a high confidence from failed predictions with a low confidence, instead of further improving the performance of anomaly detection.

\subsection{Visualizing region of diagnosis}

For the visual explanation of the decision reasoning of our CAAD model, we adopted the gradient-weighted class activation mapping (Grad-CAM) \cite{grad_cam} to "see" which regions play an important role during the inference.
Fig.~\ref{fig:CAM} shows 16 chest X-ray images from the X-VIRAL validation set, each being accompanied with the Grad-CAM maps overlaid on it. Eight cases in two top rows were diagnosed as positive by our CAAD model, while the other cases in two bottom rows were diagnosed as negative. 
It reveals that our CAAD model is able to focus on the suspected lesions and, accordingly, diagnose the input image as POSITIVE. However, if no highly suspected lesion is found, our model assigns the homogeneous activation values to almost the entire image.
The visualization demonstrates the good interpretability of our CAAD model.

\begin{table}[t]
\caption{Comparing performance of four models (w/o fine-tuning) on the X-COVID dataset. }
\vspace{+0.2cm}
\label{tab:comp_convid}
\centering
\begin{tabular}{c|c|c|c|c}
\toprule[1pt]
\multirow{2}{*}{Model} & \multicolumn{4}{c}{Performance \%} \\ \cline{2-5} 
 & Accuracy & Sensitivity & Specificity & AUC \\ \midrule[1pt]
EfficientNet \cite{tan2019efficientnet}  & 69.95 & 45.28 & 94.39 & 74.45   \\ \hline
ConfiNet \cite{corbiere2019addressing} & 68.08 & 69.81 & 66.36 & 74.89 \\ \hline
AnoDet \cite{pang2019deep}  & 73.24 & 55.66 & 90.65 & 82.97  \\ \hline
CAAD    & 72.77	& 71.70	& 73.83	& 83.61  \\ \bottomrule[1pt]
\end{tabular}
\end{table}

\subsection{Generalize to unseen X-COVID dataset}
To demonstrate its generalization ability, the well-trained CAAD model was directly tested on the unseen X-COVID dataset without fine-tuning. 
It was also compared to a binary classifier using EfficientNet \cite{tan2019efficientnet}, a binary classifier with confidence prediction (\textit{i.e.}, ConfidNet) \cite{corbiere2019addressing}, and a anomaly detection model (\textit{i.e.}, AnoDet) \cite{pang2019deep}. Note that all of these models were never trained on the COVID-19 cases.
TABLE~\ref{tab:comp_convid} gives the performance of these models on the X-COVID dataset. It reveals that 
(1) anomaly detection models are superior to both binary classifiers, especially in terms of AUC, which demonstrates the effectiveness of anomaly detection in viral pneumonia screening;
(2) confidence prediction is an effective strategy to predict failures in both a classifier and an anomaly detector, contributing to a big improvement in sensitivity; and
(3) the performance of these models, however, drops when comparing to their performance in viral pneumonia screening. 
Compared to binary classification methods, anomaly detection based AnoDet and CAAD, as one-class classification approaches, have a better ability to detect dissimilar or even previously unseen COVID-19 cases.
In summary, we expect that our model trained on the X-VIRAL dataset would have the ability to detect unseen COVID-19 cases as anomaly. Our results suggest that, despite the significant drop of sensitivity, our CAAD model still achieves an AUC of $83.61\%$ and a sensitivity of $71.70\%$ for COVID-19 screening, the highest performance obtained in our experiment.

\begin{figure}[t]
\centering
\includegraphics[height=4.9cm]{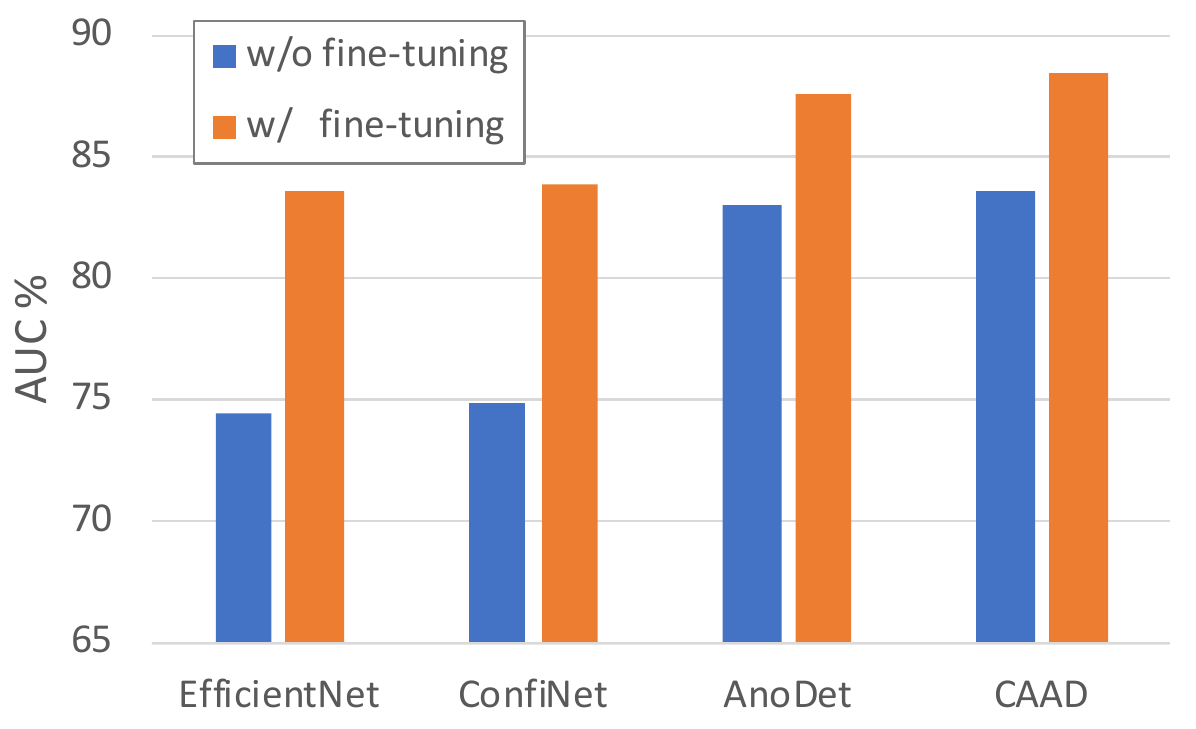}
\caption{Comparison of different models with and without using fine-tuning on the X-COVID dataset.}
\label{fig:FT}
\end{figure}

To further improve their performance on the X-COVID dataset,  we fine-tuned EfficientNet, ConfiNet, AnoDet, and our CAAD model, which had been well-trained on the X-VIRAL dataset, on the X-COVID dataset.  We conducted the two-fold cross-validation and compared the AUC values obtained by these models in Fig.~\ref{fig:FT}. It shows that fine-tuning is an effective strategy to improve the performance of all four models. Although fine-tuning narrows the gaps between the classification-based and anomaly detection-based models, it reveals that classification-based models suffer from a poorer generalization ability than anomaly detection-based models.

\begin{figure}[h]
\centering
\includegraphics[height=7.0cm]{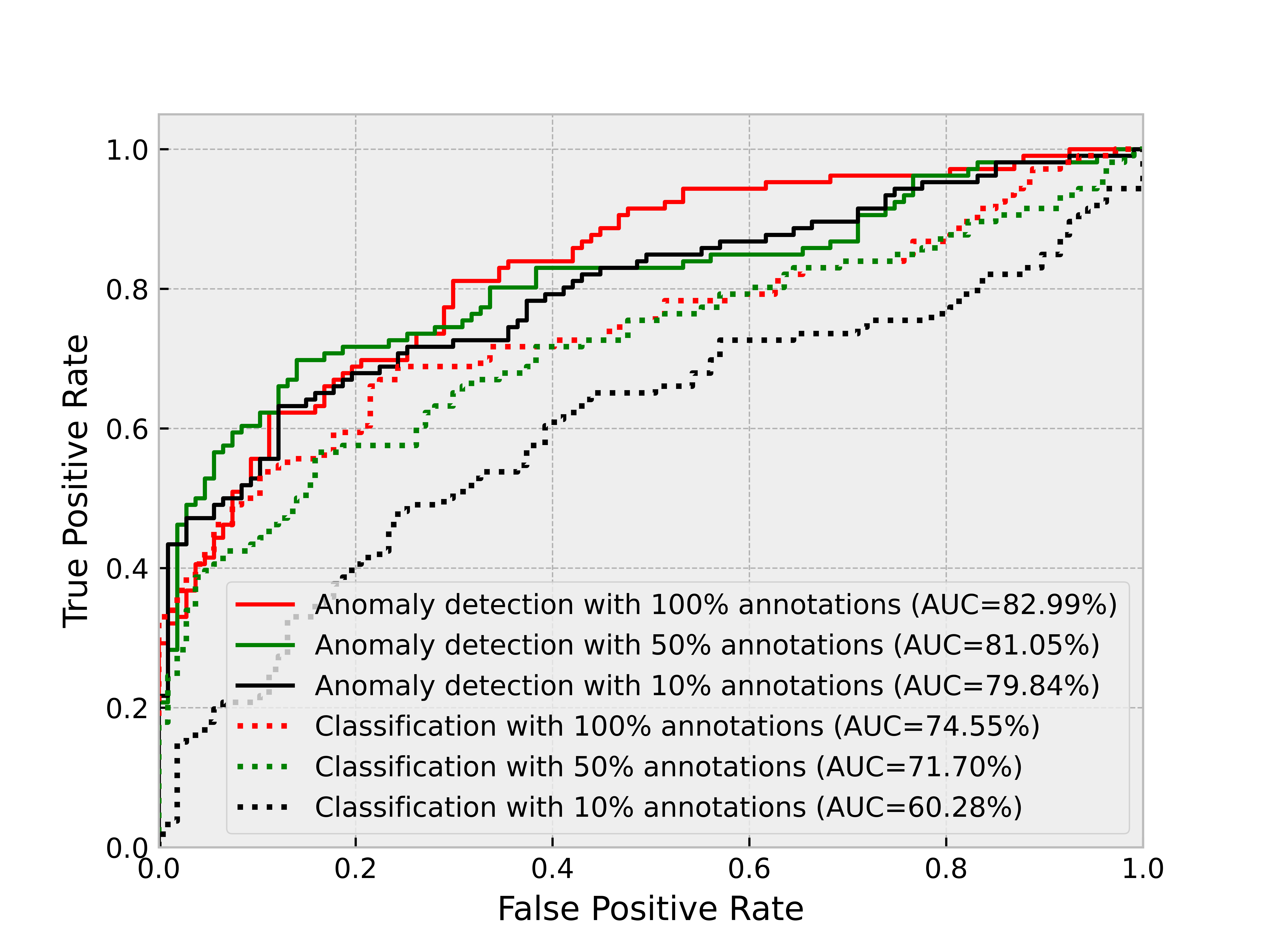}
\caption{AUC curves of anomaly detection and binary classification models obtained on X-COVID when both models were trained on X-VIRAL using all negative samples and 100\%, 50\%, and 10\% positive samples, respectively.}
\label{fig:AUC_curve}
\end{figure}

\subsection{Generalize to unseen Open-COVID dataset}
The EfficientNet, ConfiNet, AnoDet, and our CAAD model trained on the X-VIRAL dataset were further tested on the Open-COVID dataset without any fine-tuning. The performance of these models was displayed in TABLE~\ref{tab:comp_openconvid}. It shows that our CAAD model achieves an accuracy of 94.93\% for COVID-19 detection and an accuracy of 100\% for SARS and MERS detection, which are superior to those of three competing methods. Such results indicate that our CAAD model has a strong generalization ability on the unseen Open-COVID dataset.

Besides, we randomly split 2,000 negative cases (1,000 non-viral pneumonia and 1,000 healthy controls) from the X-VIRAL dataset, combined them with the healthy controls from the X-COVID dataset, and also combined the COVID-19 cases from the X-COVID and Open-COVID datasets. Thus, we have a new dataset that contains 599 positive COVID-19 cases and 2107 negative cases.
In TABLE~\ref{tab:comp_open2}, we compared the performance of two binary classification based methods ($i.e.$, EfficientNet and ConfiNet) and two anomaly detection based methods ($i.e.$, AnoDet and CAAD) on this dataset. Note that all these models never saw any COVID-19 X-ray images in the X-COVID and Open-COVID datasets. It shows that anomaly detection based models, especially our CAAD model, perform distinctly better than binary classification based models. 
Meanwhile, we also compared the performance of ConfiNet and CAAD when setting the confidence threshold $T_{conf}$ to different values. It reveals that, when we set $T_{conf}$ to $0.5$, both models can achieve not only the highest accuracy, but also high and balanced sensitivity and specificity. 
Moreover, comparing to the results in TABLE~\ref{tab:comp_open2} and TABLE~\ref{tab:comp_convid}, it shows that all four models, especially both binary classification based ones, achieve better performance on this dataset than on the X-COVID dataset. The performance gain can be largely attributed to the high accuracy of COVID-19 detection on the Open-COVID dataset, as shown in TABLE~\ref{tab:comp_openconvid}.

\begin{table}[t]
\caption{Comparing performance of four models on the Open-COVID dataset.}
\vspace{+0.2cm}
\label{tab:comp_openconvid}
\begin{center}
\begin{tabular}{c|p{1.3cm}<\centering|p{1.2cm}<\centering|p{1.2cm}<\centering|p{1.2cm}<\centering}
\toprule[1pt]
\multirow{2}{*}{Model} & \multicolumn{4}{c}{Detection accuracy (sensitivity) \%} \\ \cline{2-5} 
 & \begin{tabular}[c]{@{}c@{}}COVID \\ (493 images)\end{tabular} & \begin{tabular}[c]{@{}c@{}}SARS \\ (16 images)\end{tabular} & \begin{tabular}[c]{@{}c@{}}MERS \\ (10 images)\end{tabular} & \begin{tabular}[c]{@{}c@{}}All\\ (519 images)\end{tabular} \\ \midrule[1pt]
EfficientNet \cite{tan2019efficientnet} & 63.08 & 93.75 & 80.00 & 64.35 \\ \hline
ConfiNet \cite{corbiere2019addressing} & 90.26 & \textbf{100.00} & \textbf{100.00} & 90.75 \\ \hline
AnoDet \cite{pang2019deep} & 83.37 & \textbf{100.00} & \textbf{100.00} & 84.20 \\ \hline
CAAD & \textbf{94.93} & \textbf{100.00} & \textbf{100.00} & \textbf{95.18} \\ \bottomrule[1pt]
\end{tabular}
\end{center}
\end{table}

\begin{table}[t]
\caption{ Comparing performance of four models on the X-COVID and OpenCOVID dataset (599 positives, 2107 negatives).}
\vspace{+0.2cm}
\label{tab:comp_open2}
\centering
\begin{tabular}{c|c|c|c|c|c}
\toprule[1pt]
Method & $T_{conf}$ & Acc & Sen & Spe & AUC \\ \midrule[1pt]
EfficientNet \cite{tan2019efficientnet} & / & 75.87 & 62.6 & 79.64 & 78.92 \\ \hline
\multirow{6}{*}{ConfiNet \cite{corbiere2019addressing}} & 0.95 & 34.85 & 96.33 & 17.37 & \multirow{6}{*}{79.10} \\ \cline{2-5}
 & 0.9 & 46.45 & 92.65 & 33.32 & \\ \cline{2-5}
 & 0.8 & 57.5  & 86.14 & 49.36 & \\ \cline{2-5}
 & 0.7 & 65.56 & 82.47 & 60.75 & \\ \cline{2-5}
 & 0.6 & 70.18 & 76.63 & 68.34 & \\ \cline{2-5}
 & 0.5 & 76.39 & 63.27 & 80.11 & \\ \hline
AnoDet \cite{pang2019deep} & / & 79.79 & 69.78 & 82.63 & 83.34 \\ \hline
\multirow{6}{*}{CAAD}     & 0.95    & 42.46 & 97.33 & 26.86 & \multirow{6}{*}{84.43} \\ \cline{2-5}
 & 0.9 & 69.73 & 90.65 & 63.79 & \\ \cline{2-5}
 & 0.8 & 75.09 & 84.14 & 72.52 & \\ \cline{2-5}
 & 0.7 & 76.72 & 80.8  & 75.56 & \\ \cline{2-5}
 & 0.6 & 77.72 & 78.63 & 77.46 & \\ \cline{2-5}
 & 0.5 & 78.57 & 77.13 & 78.97 & \\ \bottomrule[1pt]
\end{tabular}
\end{table}

\subsection{Learning with less positive samples}

In the real-word scenario, it is much more difficult to collect positive (\textit{i.e.}, viral pneumonia) samples than to collect negative samples. Hence automated viral pneumonia screening has a class-imbalance nature. Anomaly detection methods avoid modeling the limited positive classes explicitly and hence better handle the class-imbalance issue than classification methods.
To verify this, we trained the anomaly detection model and binary classification model, both using EfficientNet as the feature extractor, under much more class imbalanced conditions, \textit{i.e.}, using less positive samples, and then tested both without further fine-tuning on the X-COVID dataset.
Specifically, in the training stage, we kept the number of negative samples unchanged and reduced the positive samples from 100\% to 50\% and 10\%, respectively.
The obtained AUC curves are shown in Fig.~\ref{fig:AUC_curve}. It reveals that training with less positive data leads to poor performance for both anomaly detection and binary classification models. Nevertheless, the anomaly detection model shows less performance degradation compared to the binary classification model, which indicates the advantage of using anomaly detection in terms of addressing the imbalanced problem.

\section{Discussion}
\label{Section.Discussion}

In the escalation of the COVID-19 epidemic, many attempts have been made to develop fast and accurate COVID-19 screening by means of chest medical imaging \cite{ting2020digital,changqingzhang2020tmi,zhang2020learning,shi2020largescale}.
Kang \textit{et al.} \cite{changqingzhang2020tmi} leveraged different types of features extracted from chest CT images, and introduced a multi-view representation learning method to distinguish positive COVID-19 from negative cases.
Shi \textit{et al.} \cite{shi2020largescale} presented an infection size aware random forest model to classify 1658 positive subjects confirmed COVID-19 and 1027 negative subjects without COVID-19 infection using chest CT, and achieved the sensitivity of 90.70\% and specificity of 83.30\% on this large-scale CT dataset. 
However, CT imaging takes considerably more time than X-ray imaging, and needs more complex sanitization procedures between switching patients. Besides, sufficient high-quality CT scanners may not be available in many under-developed regions, making it difficult for a timely viral pneumonia screening. In contrast, X-ray imaging is the most common and widely available chest imaging technique, playing a crucial role in clinical care and epidemiological studies \cite{cherian2005standardized, franquet2001imaging}. Most ambulatory care facilities, even in rural regions, has X-ray imaging capability. Besides, X-ray imaging is real-time which can significantly speed up the screening of a mass population in a relatively short time and at a significantly reduced cost.
Wang \textit{et al.} \cite{wang2020covid} and Apostolopoulos \textit{et al.} \cite{apostolopoulos2020covid} introduced DCNN-based binary classification models for the detection of COVID-19 cases using chest X-Ray imaging. 

Different from these COVID-19 screening works, we view COVID-19 as a novel type of viral pneumonia and attempt to distinguish it, together with other types of viral pneumonia, from non-viral pneumonia and healthy controls. 
To this end, we reformulate the binary classification problem in an anomaly detection fashion. 
Besides, we introduce a confidence prediction module to estimate the reliability of model diagnosis by learning an anomaly probability as the model confidence. 
The proposed CAAD model achieves an AUC of $83.61\%$ on COVID-19 screening, which outperforms other AI-based methods \cite{murphy2020covid}.
Although achieving a sensitivity of only 71.70\%, our CAAD model shows a screening ability that is comparable to that of radiologists, as a sensitivity of 69\% was reported in \cite{wong2020frequency}.
The reason of such a low sensitivity may attribute to the observation that some subjects have not developed radiographic visible pathology in their lungs at the early stage of viral pneumonia when the X-ray was taken \cite{weinstock2020chest}. 

\section{Conclusion}
In this paper, we have  proposed  the CAAD model for viral pneumonia screening. Our results on two chest X-ray datasets indicate that 
(1) anomaly detection works well in term of viral pneumonia screening on chest X-ray images and is superior to binary classification methods, and 
(2) learning model confidence is useful to predict failures, greatly reducing the false negatives, and
(3) our CAAD model, never seeing any COVID-19 cases, achieves an AUC of $83.61\%$ and sensitivity of $71.70\%$ on the unseen  X-COVID dataset, which is comparable to the performance of medical professionals.
Our future work will focus on further reducing the false negative rate and, if possible, decreasing the false positive rate as well. We will also investigate how to differentiate the viral pneumonia severity using chest X-ray and then detect the potentially severe cases for early interventions, which requires more clinical diagnostic information.

\bibliographystyle{IEEEtran}
\bibliography{tmi}

\end{document}